\documentclass[aop,noinfoline]{imsart} 
\usepackage{amssymb,amsmath,amsfonts}
\usepackage{mathrsfs}
\usepackage{graphicx}
\usepackage{pstricks,pst-node}
\usepackage{pst-all}
\setattribute{journal}{name}{}

\newtheorem{theorem}{Theorem}[section]
\newtheorem{proposition}[theorem]{Proposition}

\newtheorem{example}[theorem]{Example}
\newtheorem{definition}[theorem]{Definition}

\newtheorem{corollary}[theorem]{Corollary}
\newtheorem{claim}[theorem]{Claim}

\renewcommand{\Box}{{\vrule width0.6ex height1em depth0cm}}
\newenvironment{proof}{\noindent{\bf Proof:}}{\hfill \Box}

\def\build#1_#2^#3{\mathrel{\mathop{\kern 0 pt#1}\limits_{#2}^{#3}}}

\newcommand{\onetwo}{(1,2)}
\newcommand{\sat}{\mathsf{SAT}}

\newcommand{\qsat}{\mathsf{
QSAT}}

\newcommand{\onetwoqsat}{\mbox{\sf (1,2)-QSAT}}
\newcommand{\twosat}{\mbox{\sf 2-SAT}}
\newcommand{\negate}[1]{\overline{#1}}

\newcommand{\onetwoqcnf}{\mbox{\sf \onetwo-QCNF}}
\newcommand{\twocnf}{\mbox{\sf 2-CNF}}

\newcommand{\coNP}{\mathsf{coNP}}
\newcommand{\conp}{\mathsf{coNP}}

\newcommand{\calS}{{\mathcal{S}}}

\newcommand{\EE}{\mathbb{E}}

\newcommand{\pmc}{\mathbb{P}_{m,c}}

\newcommand{\pmLu}{\mathbb{P}_{m,c}}
\renewcommand{\hat}{\widehat}

\newcommand{\DELETE}[1]{}

\begin{document}

\begin{frontmatter}

\title{The threshold for random $\onetwoqsat$} \thanks{This work has been
supported by EGIDE
    10632SE, \"OAD Amad\'ee 2/2006 and ACI NIM 202. Preliminary
    versions of this article appeared in \cite{CreignouDER-08} and \cite{CreignouDER-09}}

\runtitle{The threshold for random $\onetwoqsat$}

\begin{aug}
\author{\fnms{Nadia} \snm{Creignou}},
\author{\fnms{Herv\'e} \snm{Daud\'e}},
\author{\fnms{Uwe} \snm{Egly}}
\and
\author{\fnms{Rapha\"el}  \snm{Rossignol}\corref{}\ead[label=e1]{raphael.rossignol@math.u-psud.fr}}
\runauthor{N. Creignou, H. Daud\'e, U. Egly and R. Rossignol}
\affiliation{ Universit\'{e}
   d'Aix-Marseille 2, Universit\'{e}
   d'Aix-Marseille 1, Technische Universit\"at
   Wien, and Universit\'e Paris Sud}

\address{Nadia Creignou\\
Universit\'{e} d'Aix-Marseille 2\\ Laboratoire
   d'Informatique Fondamentale\\ 163 avenue de Luminy F-13288 Marseille\\ France}
\address{ Herv\'e Daud\'e\\ Universit\'{e}
   d'Aix-Marseille 1\\ Laboratoire d'Analyse\\ Topologie et
   Probabilit\'es\\ 
   Chateau Gombert F-13453 Marseille\\ France}
\address{Uwe Egly \\ Institut f\"ur
   Informationsysteme 184/3\\ Technische Universit\"at
   Wien\\
   Favoritenstrasse 9-11\\ A-1040 Wien\\ Austria}
\address{Rapha\"el
  Rossignol \\ Universit\'{e} de
   Paris 11\\
   D\'epartement de Math\'ematiques, B\^{a}timent 425\\
   F-91405 Orsay Cedex\\ France\\
\printead{e1}}
\end{aug}

\date{\today}
\maketitle

\begin{abstract}

The $\qsat$ problem is the quantified version of the $\sat$ problem.
We show the existence of a threshold effect for the phase transition
associated with the satisfiability of random quantified extended 2-CNF
formulas. We consider boolean CNF formulas of the form $\forall X
\exists Y \varphi(X,Y)$, where $X$ has $m$ variables, $Y$ has $n$
variables and each clause in $\varphi$ has one literal from $X$ and
two from $Y$. For such formulas, we show that the threshold
phenomenon is controlled by the ratio between the number of clauses
and the number $n$ of existential variables. Then we give the exact
location of the associated critical ratio $c^{*}$.  Indeed, we prove
that $c^{*}$ is a decreasing function of $ \alpha$, where $\alpha$ is
the limiting value of $m / \log (n)$ when $n$ tends to infinity. 


\end{abstract}



\end{frontmatter}



%
%
\section{Introduction}\label{sec:introduction}
%


A significant tool for SAT research has been the study of random
instances.  It has stimulated fruitful interactions among the areas
of artificial intelligence, theoretical computer science,
mathematics and statistical physics. Recently there has been a
growth of interest in a powerful generalization of the Boolean
satisfiability, namely the satisfiability of Quantified Boolean
formulas, QBFs.  Compared to the well-known propositional
formulas, QBFs permit both universal and existential quantifiers
over Boolean variables. Thus QBFs allow  the modelling of problems
having higher complexity than SAT, ranging in the polynomial
hierarchy up to PSPACE. These problems include problems from the
areas of verification, knowledge representation and logic (see,
e.g., \cite{Egly00c}).

 Models for
generating random instances of QBF have been proposed \cite{GentW-99,ChenI-05}.
 Problems for which one can combine practical experiments with
 theoretical studies are natural candidates for first investigations
 \cite{CreignouDE-07}. In this paper, we focus on a certain subclass
 of closed quantified Boolean formulas, which can be seen as
 quantified extended $\twocnf$-formulas. These formulas bear
 similarities with $\twocnf$-formulas, whose random instances have
 been extensively studied in the literature (see, e.g.,
 \cite{ChvatalR-92, Goerdt-96, Verhoeven-99, Bollobasetal01,
   Vega-01}). At the same time, the introduction of quantifiers
 increases the complexity and requires additional parameters for the
 generation of random instances.  More precisely, we are interested in
 closed formulas in conjunctive normal form (CNF) having two quantifier
 blocks, namely in formulas of the type $\forall X \exists Y
 \varphi(X,Y)$, where $X$ and $Y$ denote distinct sets of variables,
 and $\varphi(X,Y)$ is a conjunction of 3-clauses, each of which
 containing exactly one universal literal and two existential
 ones. Evaluating the truth value of such formulas is known to be
 $\conp$-complete \cite{FloegelKKB-90}. In order to generate random
 instances we have to introduce several parameters.  The first one is
 the pair $(m,n)$ that specifies the number of variables in each
 quantifier block, i.e., in $X$ and $Y$.  The second one is $
 L=\lfloor cn\rfloor$, the number of clauses. We shall study the
 probability that a formula drawn at random uniformly out of this set
 of formulas evaluates to true as $n$ tends to infinity. We will
 denote by $\pmLu(n)$ this probability. Thus, we are interested in
$$\lim_{n\rightarrow +\infty} \pmLu(n).$$

Let us recall that the transition from satisfiability to
unsatisfiability for random $\twocnf$ formulas is sharp. Indeed, there
is a \emph{ critical value} (or a \emph{threshold}\/) of the ratio of
the number of clauses to the number of variables, above which the
likelihood of a random $\twocnf$-formula being satisfiable vanishes as
$n$ tends to infinity, and below which it goes to 1. Moreover, this
critical value is known to be $1$ (see \cite{ChvatalR-92, Goerdt-96}).

On the one hand observe that, when $m=1$, a $\onetwoqcnf$-formula
with $L$, clauses can   be seen as the conjunction of 
two independent $\twocnf$-formulas (each of which corresponds to an
assignment to the universal variable  and has on average $L/2$
clauses).  On the other hand,   when $m$ is large enough, a random
$\onetwoqcnf$-formula with  $ L=\lfloor cn\rfloor$ clauses has
essentially strictly distinct universal literals, and then behaves
as an existential $\twocnf$-formula. Thus,  we can easily prove
that the transition between satisfiability and unsatisfiability
for  random  $\onetwoqcnf$-formulas occurs when $c$ is  between 1
and 2.  Our main contribution is to identify the scale for $m$ (as
a function of $n$) at which an  intermediate and original regime
can be observed, $m=\lfloor \alpha \log n\rfloor$. Moreover,  at
this specific scale in developing further the techniques used by
Chv\`atal and Reed  \cite{ChvatalR-92}, and Goerdt
\cite{Goerdt-96},  we get the precise location of the threshold as
a function of $\alpha$.  Our main result is:

\begin{theorem}\label{thm:main}
 For any $\alpha >0$,
 there exists  $c^{*}(\alpha)>0$  such that:

  \begin{itemize}
  \item if $c<c^{*}(\alpha)$, then $\mathbb{P}_{\lfloor\alpha \ln n\rfloor,c}
\xrightarrow[n\rightarrow
    +\infty]{}
    1$, \\
  \item if $c>c^{*}(\alpha)$, then $\mathbb{P}_{\lfloor\alpha \ln n\rfloor,c}
\xrightarrow[n\rightarrow
    +\infty]{}
    0$. \\
  \end{itemize}
  Moreover, the critical ratio $c^{*}(\alpha)$   is given by 
   $$
c^{*}(\alpha) =
  \begin{cases}
  2     &
  \quad \text{if \quad $ \alpha \ln 2\leq 1$} \\
 \text{the unique root of  \quad $\displaystyle  \ln c+ \Bigl(
\frac{2}{c}-1\Bigr ) \ln(2-c) = \frac{1}{\alpha}$} &
  \quad \text{if \quad $ \alpha \ln 2>1$} \\
  \end{cases}
$$

\end{theorem}

\begin{figure}[htbp]
 \begin{center}
\psset{xunit=0.5,yunit=1.5}
\begin{pspicture}(-0.5,-0.3)(20,2.5)
 \psaxes[labels=y,ticks=all,tickstyle=top,ticksize=2pt]{->}(0,0)(-0.5,-0.3)(20,2.5)
\psecurve[linecolor=magenta]%
(-0.1,2)(0,2)(1.4,2)(1.444, 1.999874517)(2.371800000, 1.812582003)(3.299600000, 1.681618364)(4.227400000, 1.593855092)(5.155200000, 1.531000997)(6.083000000, 1.483521353)(7.010800000, 1.446188463)(7.938600000, 1.415921174)(8.866400000, 1.390787054)(9.794200000, 1.369511004)(10.72200000, 1.351216012)(11.64980000, 1.335277810)(12.57760000, 1.321239086)(13.50540000, 1.308756571)(14.43320000, 1.297567155)(15.36100000, 1.287465485)(16.28880000, 1.278288729)(17.21660000, 1.269905955)(18.14440000, 1.262210571)(19.07220000, 1.255114831)(20.00000000, 1.248545784)
  \psline[linestyle=dashed](0,1)(19,1)
  \psline[linestyle=dashed](1.444,0)(1.444,2)
   \rput(1.444,-0.2){$1/\ln 2$}
   \rput(20,-0.2){$\alpha$}
   \rput(5,-0.2){$5$}
   \rput(10,-0.2){$10$}
    \rput(15,-0.2){$15$}
 \rput(-1,2.5){$c^{*}(\alpha)$}
\label{fig:thresholdevolution}
\end{pspicture}
\caption{Evolution of the critical ratio values.}
\end{center}
\end{figure}
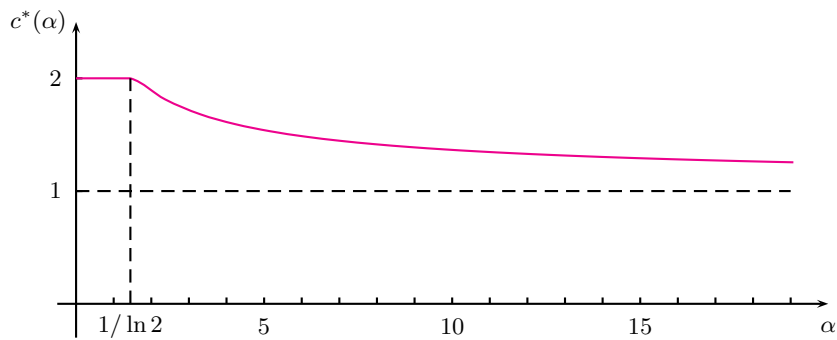

 Figure \ref{fig:thresholdevolution} shows the evolution of the critical ratio
$c^{*}(\alpha)$ as a function of $\alpha$.

The paper is organized as follows. In Section \ref{sec:complexity} we
examine the complexity of deciding the truth value of a
$\onetwoqcnf$-formula. In order to make the paper self-contained, we
give there an alternative proof of the $\coNP$-completeness of this
problem.  In Section \ref{sec:truth_char} we characterize the truth of
$\onetwoqcnf$-formulas. We introduce specific substructures,
comparable to the ones introduced by Chv\`atal and Reed in
\cite{ChvatalR-92}: we define \textit{pure bicycles}, which are
necessary to ensure the falsity of a $\onetwoqcnf$-formula, and
\textit{pure snakes}, whose appearance is sufficient to ensure the
falsity. In Section \ref{subsec:enum} we give some enumerative results
concerning pure bicycles and snakes, which will be useful for
determining the location of the threshold.  In Section
\ref{sec:transition} we present the probabilistic model and we  give first
estimates for the location of the threshold. In Section \ref{sec:main} we prove our main
result, Theorem \ref{thm:main}. Finally, Section \ref{sec:technical}
contains the proof of a technical proposition.

\section{ The complexity of  $\onetwoqsat$}\label{sec:complexity}

 A \emph{literal}\/ is a propositional variable or its negation. The
\emph{atom}\/ of a literal $l$ is the variable $p$ if $l$ is $p$ or $\negate{p}$. Literals are said to be \emph{strictly distinct } when their corresponding atoms are pairwise different.
 A \emph{clause}\/ is a finite disjunction of
literals.   A formula is in \emph{conjunctive
normal
  form}\/ (CNF) if it is a conjunction of clauses.
A formula is in $k$-CNF, if any clause consists of exactly $k$
literals.  Here we are interested in quantified propositional 
formulas of the form
$$
F=\forall X \exists Y \varphi(X,Y)
$$
where $X=\{x_1,\ldots , x_m\}$, and $Y=\{y_1,\ldots , y_n\}$, and
$\varphi(X,Y)$ is a $3$-CNF formula, with exactly one universal
and two existential literals in each clause.  We will call such
formulas \onetwoqcnf{}s. These formulas can be considered as
quantified extended $\twocnf$ formulas, because deleting the only
universal literal in each clause and removing the then superfluous
$\forall$-quantifiers result in an existentially quantified
conjunction of binary clauses. 

A truth assignment for the existential (resp. universal) variables, $X$ (resp. $Y)$ is a Boolean function $I:X\rightarrow \{0,1\}$ (resp. $Y\rightarrow \{0,1\}),$ which can be extended   to literals by  
$I(\negate{x})=1-I(x)$

A \onetwoqcnf{} formula is {\it true} (or {\it satisfiable}) if for
every assignment  to the variables $X$, there exists an assignment to
the variables $Y$ such that $\varphi$ is true under this assignment.
The exhaustive algorithm which consists in deciding whether for
all assignment to the variables $X$, there exists an assignment to
the variables $Y$ such that $\varphi$ is true provides a first
upper bound for the worst case complexity. Indeed, since the
satisfiability of a $\twocnf$ formula can be decided in linear time
\cite{AspvallPT-79}, the evaluation of the formula $\forall X
\exists Y \varphi(X,Y)$ can be performed in time $O(2^m\cdot \vert
\varphi\vert)$, where $m$ is the number of universal variables and
$\vert \varphi\vert$ denotes the size of $\varphi$. Observe that,
if $m$ is of the order of $\log n$, then it
provides a polynomial time algorithm. 

In its full generality the problem $\onetwoqsat$ is much harder as stated in the
following theorem. This theorem
was proved originally   in \cite{FloegelKKB-90}. In
order to make the paper self-contained,  we give here an
alternative proof.

 \begin{theorem} {\rm \cite{FloegelKKB-90}}\label{thm:complexity}
    The  evaluation problem $\onetwoqsat$ is $\coNP$-complete.
 \end{theorem}
 
\begin{proof}
To show membership in $\coNP$, guess a vector of truth values $v_1
,\ldots , v_m$ corresponding to $x_1,\ldots , x_m$. Replace in
$\exists Y \varphi(X,Y)$ all free occurrences of any $x_i$ by
$v_i$, remove $0$ from the clauses and delete clauses with
$1$. The resulting formula is a $2$-QCNF formula, whose
unsatisfiability (i.e. falsity) can be decided in linear time
(see \cite{AspvallPT-79} for the details).
\medskip

\noindent It remains to be shown that the problem is $\coNP$-hard.
We show this by a polynomial-time computable reduction from the
satisfiability problem for 3-CNF formulas.

Consider such a formula
$$
\alpha \colon \quad \alpha_1 \land \ldots \land  \alpha_n \quad
\quad (n\geq 2)
$$
over the variables  $\{x_2, \ldots , x_m\}$ where each $\alpha_i$ is a
disjunction of exactly three literals $l_{i,1}$, $l_{i,2}$ and
$l_{i,3}$. We construct   $\Psi(\alpha)$,   a
\onetwoqcnf\ formula.
Then we show that
\begin{equation}
  \label{eq:sat-iff-false}
  \text{$\alpha$ is satisfiable \quad\quad if and only if \quad\quad
        $\Psi(\alpha)$ is false.}
\end{equation}
The reduction is as follows. We first choose $n$ variables $y_1,
\ldots , y_n$, all of which are different from the variables $x_2,
\ldots , x_m$ occurring in $\alpha$. We take any minimally
unsatisfiable $2$-CNF formula with $n+1$ clause, e.g.,
$\psi=\bigwedge_{i=0}^n \psi_i$ where
$$
\psi_i =
  \begin{cases}
  \negate{y_1} \lor \negate{y_2}     &
  \text{if $i=0$;} \\
  y_i \lor \negate{y_{i+1}} &
  \text{if $i\in \{1, \ldots , n-1\}$;} \\
  y_{n-1} \lor y_n &
  \text{if $i=n$}.
  \end{cases}
$$

For each clause $\alpha_i= (l_{i,1} \lor l_{i,2} \lor l_{i,3})$
occurring in $\alpha$, we define
\begin{eqnarray*}
  \psi_{i,1} & = & \negate{l_{i,1}} \lor \psi_i, \\
  \psi_{i,2} & = & \negate{l_{i,2}} \lor \psi_i, \\
  \psi_{i,3} & = & \negate{l_{i,3}} \lor \psi_i.
\end{eqnarray*}

Let $x_1$ be a new variable, i.e., $x_1$ is different from the
ones in $\{y_1, \ldots , y_n\}$ and $\{x_2, \ldots , x_m\}$. Then
$$
\Psi(\alpha) \colon \quad \forall x_1 \forall x_2\cdots \forall
x_m \exists y_1 \cdots \exists y_n \; ((x_1 \lor \psi_0) \land
\bigwedge_{i=1}^n (\psi_{i,1} \land \psi_{i,2} \land \psi_{i,3})).
$$
  Obviously, the reduction is polynomial-time
computable.

\medskip

We next prove (\ref{eq:sat-iff-false}).  Observe that the formula
resulting from $\Psi(\alpha)$ by any instantiation of the $x_i$'s
is a conjunction of  clauses (maybe with repetitions) from
$\psi$. Therefore, since $\psi$ is minimally unsatisfiable, this
formula will be unsatisfiable if and only if every clause from
$\psi$ occurs.

\noindent $\Longrightarrow$: Suppose $\alpha$ is satisfiable. Take
an arbitrary truth assignment $I: X \rightarrow \{0,1\}$, which satisfies  $\alpha$.
Then, for all $i=1,\ldots , n$, there is (at least) one $j\in
\{1,2,3\}$, such that $I(l_{i,j}) = 1$. In the formula $\exists
y_1 \cdots \exists y_n \; ((x_1 \lor \psi_0) \land
\bigwedge_{i=1}^n (\psi_{i,1} \land \psi_{i,2} \land
\psi_{i,3}))$, replace all free occurrences of $x_i$ by $I(x_i)$
for $i=2,\ldots, m$ and $x_1$ by $0$. Observe that, whenever $l_{i,j}$ in $\psi_{i,j}$ (for some $j\in
\{1,2,3\}$) is true, we get $\psi_i$ after  simplification. Therefore, in the existential $2$-CNF
formula obtained after simplification it remains the clause
$\psi_0$ and at least one copy of each clause $\psi_i$ for every
$i=1,\ldots , n$ (the one resulting from $\psi_{i,j}$, for which
$I(l_{i,j}) = 1$). Therefore, this formula is unsatisfiable, thus
proving that $\Psi(\alpha)$ is false.

\noindent $\Longleftarrow$: Suppose $\Psi(\alpha)$ is false. Then,
there is a vector of truth values $v_1 ,\ldots , v_m$
corresponding to $x_1,\ldots , x_m$, such that the $2$-QCNF
formula obtained  by replacing all  occurrences of any $x_i$ by
$v_i$ is unsatisfiable. Since $\psi=\bigwedge_{i=0}^n \psi_i$ is
minimally unsatisfiable, and according to the remark above, this
means that this resulting formula contains at least one copy of
each $\psi_i$. This copy can only come from a clause $\psi_{i,j}$
for some $j\in \{1,2,3\}$. Hence, we can deduce that the
assignment $I(x_l)=v_l$ for $l=1,\ldots , m$  sets the literal
$l_{i,j}$ to true, and thus satisfies the clause $\alpha_i$.
Hence, this assignment satisfies the formula $\alpha$.
\end{proof}

\section{Truth value  of  $\onetwoqcnf$-formulas}\label{sec:truth_char}

\subsection{Pure subformulas}\label{subsec:pure}

Let us first introduce a notion of purity over sets of universal literals that will be of use to characterize the truth value  of $\onetwoqcnf$-formulas.

\begin{definition}
A (multi-)set of literals is \emph{pure} if it does not contain both a variable $x$ and its negation $\negate{x}$. 
By extension, we call a  $\onetwoqcnf$-formula, $F=\forall X \exists Y \varphi(X,Y)$, \emph{pure} if the set of universal literals occurring in $\varphi$ is pure.
\end{definition}

\begin{proposition}\label{prop:truth_char}
 A  $\onetwoqcnf$-formula is false if and only if it contains a false pure subformula.
\end{proposition}
\begin{proof}
 One direction is obvious. Suppose that the $\onetwoqcnf$-formula $F=\forall X \exists Y \varphi(X,Y)$ is false. Then, there is an assignment $I$ to the universal variables $X$ such that for all assignment to $Y$, $\varphi$ evaluates to false. Consider the subformula of $F$ obtained in keeping only the clauses for which the universal literal is assigned $0$ by $I$, and deleting the other ones. This subformula is pure (it cannot contain both a clause with a universal variable $x$ and another with $\negate{x}$ since either $x$ or $\negate{x}$ is assigned $1$ by $I$), and is false by the choice of $I$. 
\end{proof}
\medskip

Now observe that the truth value of a pure  $\onetwoqcnf$-formula $F$ is the same as the truth value of the existential $\twocnf$ formula $F_Y$ obtained in removing the universal literal in each clause and then deleting the universal quantifiers. Therefore, we can appeal to the work of Chv\`atal and Reed \cite{ChvatalR-92} in order to identify substructures that are sufficient (respectively, necessary) to ensure falsity.
On the one hand Chv\`atal and Reed exhibited elementary unsatisfiable $\twocnf$-formulas, called \emph{snakes}. On the other hand they identified extremal substructures, called \emph{bicycles}, that appear  in any unsatisfiable $\twocnf$-formula. Thus, we can define \emph{ pure snakes} and \emph{pure bicycles}.
\begin{definition}
  A pure \emph{snake}  of length $s+1\geq 4$, with $s+1=2t$,  is a set
  of  $s+1$ clauses $C_0,\ldots, C_s$ which have the following
  structure: there is a sequence of  
 $s$
strictly distinct existential literals $w_1, \ldots, w_s$, and a pure sequence of $s+1$ universal literals $v_0, \ldots, v_s$  such
that, for every $0\le r\le s$, $C_r=(v_r\lor \negate{w_r}\lor w_{r+1})$ with $w_0=w_{s+1}=\negate{w_t}$.
\end{definition}

\begin{definition}
  A \emph{ pure bicycle} of length $s+1\geq 3$, is a set of  $s+1$ clauses $C_0,\ldots, C_s$ which have the following structure:
there is a sequence of  $s$
strictly distinct existential literals $w_1, \ldots, w_s$, and a pure sequence of $s+1$ universal literals $v_0, \ldots, v_s$  such
that, for  $0 < r <  s$, $C_r=(v_r\lor \negate{w_r}\lor w_{r+1})$, $C_0=(v_0\lor u\lor w_{1})$ and $C_s=(v_s\lor \negate{w_s}\lor v)$  with literals $u$ and $v$ chosen from $w_1,
\ldots, w_s, \negate{w_1}, \ldots,\negate{w_s}$ with $(u,v)\not
=(\negate{w}_s,w_1)$. 
\end{definition}

Thus, we get the following proposition.
\begin{proposition}\label{prop:certificates}\ 

 \begin{itemize}
 \item Every $\onetwoqcnf$-formula that contains a pure snake is false.
\item Every   $\onetwoqcnf$-formula that is false, contains a pure bicycle.
\end{itemize}
\end{proposition}

\subsection{Enumerative results}\label{subsec:enum}

\begin{proposition}\label{prop:uppercertificatespure}
 Let $m$ be the number of universal variables and let $n$ be the number of existential variables we can choose from. 
  
   \begin{itemize} 
  \item The number of snakes of length $s+1$ is 
\begin{equation}\label{eqn:nbSnakes}
   (n)_s2^s\, d(m,s+1)\;
      \end{equation}
    where
  \begin{equation}\label{eqn:cms}
    d(m,s+1)=\sum_{k=1}^{min(m,s+1)} {m\choose k}\cdot 2^k\cdot \calS(s+1,k)
    \cdot k!\;
 \end{equation}
  with $\calS (m,k)$ denoting the Stirling number of the second kind,
  and 
  
  $(n)_s = (n-1)\cdots (n-s+1)$.

\item Given a pure snake $A_0$ of length $s+1=2t$. For every $1\leq i
  \leq 2t-1$,  let $N_{m,s}(i)$ denote the number  of pure snakes $B$ of
  length $s+1$ such that $A_0$ and $B$ share exactly  $i$ clauses.  Then for  $1\leq i \le t-1$ 

\begin{equation}\label{majo1N}  
N_{m,s}(i)\leq 2(s+1)^3 \Biggl [ \sum_{h=1}^{2t} \bigl ( \frac{ (s+1)^3}{n-s}\bigr )^h \Biggr ]  (n)_{s-i}2^{s-i}\, d(m,s+1-i)
\end{equation} and for  $t\leq i \leq 2t -1$
\begin{equation}\label{majo2N}
 N_{m,s}(i)\leq 2 (s+1)^3 \Biggl [ \sum_{h=0}^{2t} \bigl ( \frac{ (s+1)^3}{n-s}\bigr )^h \Biggr ] (n)_{s-i}2^{s-i}\, d(m,s+1-i)
\end{equation}
hold.

\item The number of bicycles of length $s+1$   is
\begin{equation}\label{eqn:nbBicycles}
   [(2s)^2 -1] (n)_s2^s\, d(m,s+1)\;
  \end{equation}
\end{itemize}

\end{proposition}
\begin{proof}
Given a literal $w$, let $\vert w\vert$ denote its underlying variable. Observe that a snake of length $s+1=2t$ contains $s$ distinct variables.  Moreover, every variable $\vert w_i\vert$ appearing in a snake occurs exactly twice (once positively and once negatively), except for $\vert w_0\vert$ which occurs four times (twice positively and twice negatively). This special variable will be called the \emph{double point} of the snake. A snake can be   described by a (circular) sequence of existential literals $w_0, w_1,\ldots w_s (w_0)$ (with $w_0=\negate{w_t}$), together with the corresponding pure sequence of universal literals $v_0, v_1,\ldots v_s$. 

Choosing a snake  of length $s+1$ comes down to choose a sequence of  $s$ strictly distinct
literals $w_{1},\ldots
  ,w_{s}$,   and then   choose the pure sequence of $s+1$ universal literals $v_0, \ldots,
  v_s$ (they are not necessarily distinct but no literal can be the
  complement of another).
  Let $d(m,s+1)$ be the number of pure sequences of literals of length
  $s+1$, having a set of $m$ variables from which the literals can be
  built.  Let us recall that $\calS (m,k)\cdot k!$ is the number of
  applications from a set of $m$ elements onto a set of $k$ elements.
  A pure sequence of literals of length $s+1$ is obtained by exactly
  one sequence of choices of the following choosing process.
  \begin{enumerate}
  \item Choose the number $k$ of different variables occurring in the
    sequence.
  \item Choose the $k$ variables.
  \item For each such variable, choose whether it occurs positively or
    negatively.
  \item Choose their places in the sequence.
  \end{enumerate}
This gives the announced number of snakes.
\medskip

Given a pure snake $A_0$ of length $s+1=2t$. Let $N_{m,s}(i)$ be the
number of pure snakes $B$ of length $s+1$  such  that $A_0$ and $B$ share exactly  $i$
clauses. If $i\leq 2t-1$, this number can be decomposed as
$$N_{m,s}(i)=\sum_{j\ge i+1} N_{m,s}(i,j)$$ where $N_{m,s}(i,j)$ is the number of  pure snakes $B$ such  that $A_0$ and $B$ share exactly  $i$ clauses and $j$ variables. In the rest of the proof, for more readability we omit the subscripts $m,s$ in $N_{m,s}(i,j)$, thus writing $N(i,j)$. Now we are looking for  upper bounds on the $N(i,j)$.

Let us note that the intersection of $A_0$ and $B$ can be read on the (circular) sequence of literals
 $w_0, w_1,\ldots w_t, \ldots w_s (w_0)$, where $w_t=\negate{w_0}$.  In order to get $i$ clauses and $j$ variables in common, one has to choose $k=(j-i)$ blocks of consecutive literals in this sequence. 
We make a case distinction according to whether the two snakes  $A_0$ and $B$ have the same double point or not.
\begin{itemize}
 \item $N^{a}(i,j)$ denotes the  number of pure snakes $B$ of length $s+1$ such
that  $A_0$ and $B$ share exactly  $i$ clauses and $j$ variables, and have the same double point $\vert w_0\vert$,
\item $N^{b}(i,j)$ denotes the  number of pure snakes $B$ of length $s+1$ such
that  $A_0$ and $B$ share exactly  $i$ clauses and $j$ variables, and do not have    the same double point.\\
\end{itemize}
Thus  $N(i,j)=N^{a}(i,j)+N^{b}(i,j).$ 

Let us  first consider $N^{a}(i,j).$
Observe that in the special case when $j=i+1$ (only one block), and $A_0$ and $B$ have the same double point, then $i$ is necessarily equal to or larger than $t$. Therefore,  
\begin{equation}\label{eqn:boundN0}
\hbox{ for } 1\le i\le t-1, \quad N^{a}(i,i+1)=0\ .
\end{equation} 
In the general case, to count $N^{a}(i,j)$, we
perform the following sequence of choices :

\begin{tabular}{ll}
$(i)$&  the intersection $A_0\cap B$ such that it has $i$ clauses and $j$ variables,\\
$(ii)$& the sequence of strictly distinct existential literals that are  in $B\setminus (A_0\cap B)$\\
$(iii)$&  the places of the $k$ blocks of $A_0\cap
B$ among the literals chosen in $(ii)$,\\
$(iv)$&  the universal literals occurring in  the clauses of $B\setminus (A_0\cap B)$.
\end{tabular}

\emph{Step (i)}. To build the intersection $A_0\cap B$, we choose $2k$
literals  in the sequence representing  $A_0$. They represent the first and last literals of the
$k$ blocks of $A_0\cap B$. The first literal is chosen after or at $\omega_0$. 
To
define completely the intersection, we need to know whether this first
literal is the beginning or the end of a block, so we get
at most $2\binom{s+1}{2k}\le (s+1)^{2k}$ possible choices.

\emph{Step (ii)}. Notice that   $\vert w_0\vert $ is the double point of $B$. So, it remains only to choose a sequence of $s-(j-1)$  strictly distinct literals.
Thus,
we have at most $(n)_{s+1-j}2^{s+1-j}$ possible choices.

\emph{Step (iii)}.
 We need to choose how the $k$  blocks will be plugged among the ``remaining literals''  chosen in
$(ii)$. 
This
leads to at most $(s+1)^{k}$ possible choices.

\emph{Step (iv)}. There are $s+1-i$ universal literals to choose, and they must be chosen in
a pure way. So, there are at most $d(m, s+1-i)$ choices.

Thus, since   $k=j-i$ we obtain that   for $1\leq i \leq 2t-1, \ j \geq i+1$
\begin{equation}\label{eqn:boundN1}
 N^{a}(i,j) \leq (n-s) \Bigl ( \frac{ (s+1)^3}{n-s} \Bigr ) ^{j-i}  (n)_{s-i}2^{s-i}\, d(m,s+1-i)\, .
\end{equation}
The enumeration of $N^{b}(i,j)$ differs from the one of $N^{a}(i,j)$ only at step (ii). Indeed, when $B$ does not have $\vert w_0 \vert $ as a double point, at step (ii) we have first to choose  a sequence of $s-j$  strictly distinct literals (thus having  determined the $s$ variables occurring in $B$), and then choose one of these $s$   variables as the double point. Hence, we have at most $s(n)_{s-j}2^{s-j}$ choices.
Thus, we get  for  $1\leq i \leq 2t-1$ and $ j \geq i+1$
\begin{equation}\label{eqn:boundN2}
   N^{b}(i,j) \leq s \Bigl ( \frac{ (s+1)^3}{n-s} \Bigr ) ^{j-i} (n)_{s-i}2^{s-i}\, d(m,s+1-i)\, .
\end{equation}
Then, equation  $(\ref{majo1N})$ follows from (\ref{eqn:boundN0}), (\ref{eqn:boundN1})and  (\ref{eqn:boundN2}) while  $(\ref{majo2N})$  follows from (\ref{eqn:boundN1}) and (\ref{eqn:boundN2}).\\
 
 The enumeration of bicycles is similar to the one of snakes. We just have to choose in addition $u$ and $v$ among $w_1,\ldots, w_s,\negate{w_1},\ldots, \negate{w_s}$ such that   $(u,v)\ne (\negate{w_s}, w_1)$. This explains  the  extra factor, $ [(2s)^2 -1]$, in (\ref{eqn:nbBicycles})

\end{proof}

\section{Location of the   transition for
$\onetwoqsat$}\label{sec:transition}

We consider formulas  built on $n$ universal variables and $m$ existential variables. Thus we have 
$\displaystyle N= m  {n\choose 2}  2^3= 4   m
 n(n-1)$ different clauses at hand.  
 We may  establish our result in considering random formulas obtained by taking
each one of the $N$ possible clauses
  independently from the others with probability $p\in]0,1[$. Let $c>0$, 
it is well known, see for instance  \cite[Sections 1.4 and 1.5]{JansonLR-00},  that the threshold obtained in this model translates to the
model alluded to in the introduction -- in which $ L=\lfloor cn\rfloor$, distinct clauses are picked
uniformly at random among all the $N$ possible choices --,  when $p=\frac{L}{4mn(n-1)}$.
Thus, from now on we shall always suppose that
$p=\frac{c}{4mn}$, and we continue to denote by $\pmc(n)$ the probability
that a random formula in this model is satisfiable. We are
interested in studying $\displaystyle \lim_{n\rightarrow +\infty }\pmc(n)$
as a function of the parameters $m$ and $c$. Any value of $c$ such
that $\pmc(n) \to 1$ (resp.\ such that $\pmc(n) \to 0)$ gives a
lower (resp.\ upper) bound for the threshold effect associated to
the phase transition.\\

 Let us recall that the $\twosat$ property exhibits a sharp transition, with a
critical value equal to 1 (see \cite{ChvatalR-92} and \cite{Goerdt-96}). From
this result it is easy to deduce that the phase transition from satisfiability
to unsatisfiability for  
 \onetwoqcnf\ formulas occurs
when $1\le c\le 2$. 
 
\begin{proposition}\label{prop:first_estimates}
  Let $m=m(n)$ be any sequence of integers.
  \begin{itemize}
  \item If $c<1$ then $\pmc(n)\xrightarrow[n\rightarrow \infty]{}
    1$.\\
  \item If $c>2$ then $\pmc(n)\xrightarrow[n\rightarrow \infty]{} 0$.
  \end{itemize}

\end{proposition}
\begin{proof}
Let $F$ be  a random $\onetwoqcnf$-formula.
 Let us consider $F_t$, the
  $\twocnf$ formula obtained from $F$ by setting all the variables
  $x_1,\ldots ,x_m$ to \emph{true} and omitting all quantifiers. If
  $F$ is satisfiable, then so is $F_t$. Notice that $F_t$ can be
  obtained by picking independently each possible 2-clause with
  probability
  $$q(n)=1-(1-p(n))^m=\frac{c}{4n}+O\left(\frac{1}{n^2}\right).$$ Thus
  the average number of clauses in $F_t$ is equal to $$4{n\choose
    2}\cdot q\sim c/2\cdot n.$$  It follows from the threshold of 2-SAT
  \cite{ChvatalR-92,Goerdt-96} that $F_t$ is unsatisfiable with
  probability tending to 1 if $c>2$. Thus, the same holds for $F$.

  Now, we look at the existential part of the formula, $F_Y$. Observe
  that if $F_Y$ is satisfiable, then so is $F$.  In $F_Y$, each of the
  $4\binom{n}{2}$ 2-clauses appears independently with probability
  $$q'(n)=1-(1-p(n))^{2m}=\frac{c}{2n}+O\left(\frac{1}{n^2}\right).$$
  Therefore,
  the threshold of 2-SAT tells us that when $c<1$, the formula $F_Y$
  is satisfiable with probability tending to one. The same holds
  for $F$.
\end{proof}
\medskip

\section{Proof of the main result} \label{sec:main}

\subsection{General inequalities}
Let $B_s$ and $X_s$ be respectively the
number of pure bicycles and pure snakes of length $s+1$   in a random
\onetwoqcnf  \ formula.     Les us recall that in such a formula, each
clause is chosen with probability $\displaystyle
p=\frac{c}{4mn}$. Hence, if  $\EE_{m,c}(B_s)$ and $\EE_{m,c}(X_s)$
denote   the average number of  bicycles and snakes of length $s+1$ in
a random \onetwoqcnf  \ formula,    we get from (\ref{eqn:nbSnakes}),
(\ref{eqn:cms}) and (\ref{eqn:nbBicycles}) the following two equations:
    \begin{equation} \label{MeanSnakes}
    \EE_{m,c}(X_s) =  p^{s+1}(n)_s2^sd(m,s+1) 
    \end{equation}
    \begin{equation}\label{Meanbicycles}
      \EE_{m,c}(B_s)= \EE_{m,c}(X_s) ((2s)^2 -1). 
      \end{equation}
    In order to prove that $c^{*}$ is the critical value for the (decreasing) satisfiability  property for  $\onetwoqcnf$-formulas, we will use  two  sequences of inequalities. The first one follows from    Proposition \ref{prop:certificates} and Markov
inequality applied on the number of bicycles. We have 
\begin{equation}\label{eqn:first_moment}
  1-\pmc(n)\le \Pr \Bigl (\sum_{s\geq 2} B_s \geq 1\Bigr )\le \sum_{s\geq 2} \EE_{m,c}(B_s).
\end{equation}
The second one is obtained in   considering  the number of snakes. Proposition
\ref{prop:certificates} and  a general exponential inequality given in
\cite[Theorem 2.18  ii)]{JansonLR-00} show that   for any $s\geq 3$ 
\begin{equation}
\label{eqn:Janson}
  \pmc(n)\le \Pr(X_s=0)\le \exp \Biggl  ( - \frac{\EE_{m,c}(X_s)}{1+ \sum_{i=1}^{s} N_{m,s}(i) p^{s+1-i}} \Biggr )
\end{equation}

Finally,  recall that we can  suppose that $1<c<2$, according to  Proposition \ref{prop:first_estimates}. 

\subsection{When the critical ratio is equal to $2$}\label{subsec:c=2}

Let us start with a proposition which enables to control the mean
number of bicycles for any $c$ in $]1,2[$.

\begin{proposition}\label{prop:bicyclebound}
 For any $1<c<2$, the following statements hold when  $n$ tends to  infinity 
\begin{itemize}
\item 
if  $\displaystyle m\leq\frac{\ln n}{\ln 2}$ then  
$\displaystyle \sum_{s\geq 2}\EE_{m,c} (B_s) = o(1)$

\item  if $m=\lfloor \alpha \ln n\rfloor$  with $\alpha \ln 2 >1$ then $\displaystyle
  \sum_{s\ge 2\frac{\alpha \ln 2 -1}{\ln 2 -\ln c} \ln n }\EE_{m,c} (B_s) = o(1).$  
\end{itemize}
\end{proposition}
\begin{proof}
Let us recall that the coefficient $d(m,s+1)$ occurring in $\EE_{m,c} (B_s)$ is  the number of pure sequences of literals of length
  $s+1$, when we have   $m$ variables from which the literals can be
  built.   Note that $d(m,s+1)$ is bounded
from above by $2^{\min\{m,s+1\}}$ times the number of applications
from $\{1,\ldots,s+1\}$ to $\{1,\ldots,m\}$. Therefore,
\begin{equation}
  \label{eq:majostirlingtriviale} d(m,s+1)\leq
  2^{\min\{m,s+1\}}m^{s+1}.
\end{equation}
From (\ref{Meanbicycles}),  it follows that  if   $s<m$ then  $\displaystyle  \EE_{m,c} (B_s) \leq \frac{c^{s+1} s^2 }{n}$. Thus 
\begin{equation}\label{Majsnake1}
\sum_{s<m} \EE_{m,c}(B_s)    \leq \Bigl ( \frac{c}{c-1} \Bigr ) \ m^2 \, \frac{c^m}{n}.
 \end{equation}
If $s\ge m$,  then (\ref{eq:majostirlingtriviale}) gives $\displaystyle \EE_{m,c} (B_s) \leq \Bigl (\frac{c}{2}\Bigr )^{s+1} s^2  \frac{2^m }{n}.$ When $0<x<1$ and $r\geq 2$, standard computations show
that
\begin{equation}
  \label{eqn:majosumcarree} \sum_{s=r}^{\infty}s^2x^s\leq
  r^2\frac{x^r}{(1-x)^3}.
\end{equation}
Hence  we get 
\begin{equation}\label{Majsnake2} 
 \sum_{s\ge r} \EE_{m,c}(B_s)   \leq\frac{c2^mr^2\left(\frac{c}{2}\right)^r}{n(1-c/2)^3} . 
\end{equation}
The proof  of  Proposition \ref{prop:bicyclebound}  is now an easy consequence of  (\ref{Majsnake1}) and (\ref{Majsnake2}). 
 \end{proof}


 Theorem \ref{thm:main} when $\alpha\ln 2\leq 1$ follows from Proposition
 \ref{prop:bicyclebound}, inequality (\ref{eqn:first_moment})  and
 Proposition \ref{prop:first_estimates}.

In the sequel, we consider the case where $m= \lfloor \alpha\ln n\rfloor$, with $\alpha>1/\ln 2$. 

\subsection{The critical ratio as a function of $\alpha$}

 The main  difficulty when dealing with $\EE_{m,c}(B_s)$ and $\EE_{m,c}(X_s)$ is to handle the coefficient $d(m,s+1)$ given in Proposition \ref{prop:uppercertificatespure} 
$$ d(m,s+1) =\sum_{k=1}^{min(m,s+1)} {m\choose k}\cdot 2^k\cdot \calS(s+1,k)
    \cdot k!\ .$$
    
    First, let us denote for  $1\le k \le \min(m,s+1)$ 
         \begin{equation}\label{keyexpression}
     G_{m,c}(k,s+1)= 2^s \,(n)_s \,{m\choose k} \,2^k \,\calS (s+1,k) \,k! \,\Bigl ( \frac {c}{ 4mn}\Bigr )^{s+1}.
     \end{equation}
From (\ref{MeanSnakes}) and (\ref{Meanbicycles}), the behavior of $\EE_{m,c}(X_s)$ and $\EE_{m,c}(B_s)$ is clearly governed by the  coefficients $G_{m,c}(k,s+1)$. Indeed,  since $\displaystyle p=\frac{c}{4mn}$ we get      
         \begin{equation}\label{keyrelation}
\EE_{m,c}(B_s)=  \sum_{k=1}^{\min(m,s+1)} G_{m,c}(k,s+1)((2s)^2 -1) = ((2s)^2 -1) \EE_{m,c}(X_s)
\end{equation}

Second, we will need   better bounds than the one given in (\ref{eq:majostirlingtriviale}). We will use  well-known estimates for binomial coefficients.  If $1\le b \le a$, then the following inequalities hold:
\begin{equation}\label{binomialbounds}  \sqrt{\frac{1}{a}}\left (\frac{a}{b}\right )^b\cdot \left
    (\frac{a}{a-b}\right )^{a-b} \le  {\binom{a}{b}}\le \left (\frac{a}{b}\right )^b\cdot \left
    (\frac{a}{a-b}\right )^{a-b} \ .
   \end{equation}
Then, from  \cite{Temme-93}, we have the following bounds
for Stirling numbers of the second kind. There exist  
$K>0$ and $K'>0$ such that, for $1\le b\le a$, the following inequalities hold:
\begin{equation}\label{Stirlingbounds}
K  \sqrt{\frac{b}{a}}\left (\frac{e^{x_0}-1}{x_0}\right
)^b \left ( \frac{a}{e}\right )^a x_0^{b-a}\le b! \calS(a,b)\le K'   \sqrt{b}\left (\frac{e^{x_0}-1}{x_0}\right
)^b \left ( \frac{a}{e}\right )^a  x_0^{b-a} \end{equation} 
where $x_0>0$ is a function of $b/a$ defined implicitly for $b<a$ by
 $1-e^{-x_0}=\frac{b}{a}x_0$, and for $a=b$ by $x_0=0$.
The conventions are that $0^0=1$ and $\frac{e^0-1}{0}=1$.

By using these precise results, already used in \cite{DuboisB-97} and \cite{CreignouDE-07},  it appears that the behaviour of the coefficients $G_{m,c}(k,s+1)$ and so the one of the average number of snakes or bicycles, is  governed by a continuous function of several real variables. From (\ref{keyexpression}), (\ref{binomialbounds}) and (\ref{Stirlingbounds})  we obtain: 
\begin{proposition}\label{thebig} There exist $A>0$ and $B>0$ such that   for any  $c>0$, for every positive  integers $n, m, s$ and $k$ such that   $k \le \min(m,s+1)$  :  
  \begin{equation}\label{minMaj}
 \frac{A\, (n)_s\sqrt{k}}{ n^s\sqrt{m (s+1)}}\  n^{g_{\frac{m}{\ln n},c}(\frac{k}{\ln n},\frac{s+1}{\ln n})}\le G_{m,c}(k,s+1)\leq B \sqrt{m}\  n^{g_{\frac{m}{\ln n},c}(\frac{k}{\ln n},\frac{s+1}{\ln n})}
 \end{equation}
 where $g_{\alpha,c}$ is the continuous  function  on  $\mathcal{D}_{\alpha}= \{ (\beta,
 \gamma) \mid 0< \beta \le \alpha  \hbox { and } \beta \le \gamma\}$ defined for $0<\beta < \gamma$ by
 \begin{equation}\label{eqn:def-of-g}
   g_{\alpha,c}(\beta, \gamma)=\ln \left\lbrack\frac{1}{e}\left (
    \frac{c\gamma}{2ex_0\alpha}\right ) ^\gamma\cdot
  \frac{\alpha^\alpha}{\beta ^\beta(\alpha-\beta)^{\alpha-\beta}}
  \cdot 2^\beta\cdot (e^{x_0}-1)^\beta\right\rbrack,
\end{equation}
 with $\displaystyle 1-e^{-x_0}=\frac{\beta}{\gamma}x_0$ and  $ \displaystyle g_{\alpha,c}(\beta, \beta) = \ln \left\lbrack\frac{1}{e}\left (
    \frac{c}{e\alpha}\right ) ^\beta\cdot
  \frac{\alpha^\alpha}{(\alpha-\beta)^{\alpha-\beta}}
  \right\rbrack.$
  \end{proposition}

  Recall that we have taken $m= \lfloor \alpha\ln n\rfloor$. Observe
  that the second part of Proposition \ref{prop:bicyclebound} together with
  (\ref{Meanbicycles}) indicates that  long  snakes, and similarly
  long bicycles, of length $\gg \ln n$, have asymptotically no chance
  to appear when $\alpha>1/\ln 2$ and $c\in]1,2[$.  Therefore, in our study we will  focus  on snakes  of length proportional to $\ln n$.    Hence, let us
set $\beta=k/\ln n$, $\gamma=(s+1)/\ln n$. The following result will point out for each $\alpha$, the values of $k$ and $s$ that contribute the most to the average number. Indeed we  will prove the following central result :
\begin{proposition}\label{prop:function g}
 Let  $1<c<2$, and  for any $\alpha$ let $\mathcal{D}_{\alpha}$ be the following domain   
 $$\mathcal{D}_{\alpha} = \{ (\beta,
 \gamma) \mid  0< \beta \le \alpha  \hbox { and } \beta \le \gamma\}.$$
 The function $g_{\alpha,c}$ 
 defined by (\ref{eqn:def-of-g})  has a  global maximum on $\mathcal{D}_{\alpha}$, given by its unique stationarity point in $\mathcal{D}_{\alpha}$. More precisely 
\begin{equation}\label{Themax}
\max _{\mathcal{D}_{\alpha}}g_{\alpha,c}(\beta, \gamma)= g_{\alpha,c}(\hat\beta(\alpha,c), \hat\gamma(\alpha,c))= \alpha H(c)-1
\end{equation}
with
$\displaystyle \hat\beta= \frac{2\alpha(c-1)}{c},$
    $\displaystyle\hat\gamma=\frac{-2\alpha\ln (2-c)}{c},$ \ 
$\displaystyle H(c)=\ln c+ \Bigl( \frac{2}{c}-1\Bigr ) \ln(2-c).$

Moreover, for any domain $V_{\alpha} \subset   \mathcal{D}_{\alpha}$ such that $(\hat\beta, \hat\gamma) \notin \overline{V_{\alpha} }$ then 
\begin{equation}\label{Thepasmax}
\max _{V_{\alpha} }g_{\alpha,c}(\beta, \gamma)< \alpha H(c)-1\ .
\end{equation}

\end{proposition}
The proof of this result is rather technical, so we postpone it to the next section.\\

Now we can prove  Theorem \ref{thm:main} when $\alpha\ln 2>1$. In other words that,  when  
$ \alpha \ln 2>1$, the critical ratio $c^{*}(\alpha)$ is  the unique root of  $\alpha \ H(c)=1.$ For this,  we will use  two  corollaries of Proposition \ref{thebig} and Proposition \ref{prop:function g}.
\begin{corollary}\label{lowerbound}
Let $\alpha>1/\ln 2$ and $c< 2$ be such that $\alpha H(c)<1$.  Then, as $n$ tends to infinity 
$$\sum_{s\ge 2} \EE_{\lfloor \alpha \ln n \rfloor, c} (B_s) = o(1).$$ 
\end{corollary}
\begin{proof} From Proposition \ref {prop:bicyclebound}, we have $\displaystyle \sum_{s\ge 2 \frac{\alpha \ln 2 -1}{\ln 2 -\ln c} \ln n }\EE_{\lfloor \alpha \ln n \rfloor, c} (B_s) = o(1).$ Then, from (\ref {keyrelation}), the upper bound (\ref {minMaj}) and (\ref{Themax}) we get 
 $$\sum_{s< 2 \frac{\alpha \ln 2 -1}{\ln 2 -\ln c} \ln n }\EE_{\lfloor
   \alpha \ln n \rfloor, c} (B_s) \le B'\,  \alpha^{3/2} (\ln n)^{9/2}
 \Bigl [\frac{\alpha \ln 2 -1}{\ln 2 -\ln c}\Bigr ]^3 n^{\theta_n}\;,$$
 with $\theta_n \to g_{\alpha, c}(\hat{\beta}, \hat{\gamma})=\alpha H(c)  -1<0$. Therefore,
  $\displaystyle \sum_{s< 2 \frac{\alpha \ln 2 -1}{\ln 2 -\ln c} \ln n }\EE_{\lfloor \alpha \ln n \rfloor, c} (B_s) = o(1).$
\end{proof}\medskip

With (\ref{eqn:first_moment}), this corollary  proves that,  when $\alpha \ln 2>1$ and   for any $c<c^{*}(\alpha)$, we have  $\pmc(n)=1-o(1)$. \smallskip

In considering (\ref{eqn:Janson}) with $s+1=\lfloor\hat{\gamma} \ln n
\rfloor= 2t$,  it will follow easily from the  corollary  given below
that, when $\alpha \ln 2>1$,  for any $c>c^{*}(\alpha)$ we have
$\pmc(n)=o(1)$. This will end the proof of Theorem \ref{thm:main}. Note that the coefficients $N_{m,s}(i)$ appearing in the following corollary are the ones defined in Proposition \ref{prop:uppercertificatespure}.
\begin{corollary}\label{upperbound}
Let $\alpha>1/\ln 2$ and $c< 2$ be such that $\alpha H(c)>1$, and let
$s+1=\lfloor\hat{\gamma} \ln n \rfloor$. Then there exist
$0<\delta<2(\alpha H(c)-1)$, $C>0$ and $D>0$ such that   

$$ \EE_{\lfloor \alpha \ln n \rfloor, c} (X_s) \ge C \ n^{\alpha H(c) -1 - \frac{\delta}{2}}$$
and 
$$\sum_{i=1}^s N_{\lfloor \alpha \ln n \rfloor,s}  (i) \Bigl ( \frac{c}{4mn} \Bigr )^{s+1-i} \le D \   n^{\alpha H(c) -1 - \frac{2\delta}{3}}.$$
\end{corollary}
\begin{proof} 
From  (\ref{Thepasmax}) in Proposition \ref{prop:function g}, we first choose $\delta\in ]0,2(\alpha H(c)-1)[$ such that 
 $$ \max_{\{ (\beta, \gamma) s.t. \gamma < \frac{\hat\gamma}{2}\} \cap \mathcal{D}_{\alpha}} g_{\alpha, c}(\beta, \gamma)\le  \max_{\mathcal{D}_{\alpha}} g_{\alpha, c}(\beta, \gamma)- \delta.$$
Again in using (\ref {keyrelation}) and   the lower bound in (\ref{minMaj}), we can find $C>0$ such that for $s+1=\lfloor\hat{\gamma} \ln n\rfloor$ 
$$\EE_{\lfloor \alpha \ln n \rfloor, c} (X_s) \ge C \ n^{ g_{\alpha, c}(\hat{\beta}, \hat{\gamma})-\frac{\delta}{2}}.$$
As $ g_{\alpha, c}(\hat{\beta}, \hat{\gamma})=\alpha H(c) -1 $, the first assertion is proved.

Then, with $\displaystyle p= \frac{c}{4mn}$,   from (\ref{majo1N}) and  (\ref{majo2N})  we get first for $1\le i<t$ 
$$N_{m,s}  (i) \ p^{s+1-i} \le 
2 (s+1)^3 \sum_{k=1}^{\min (m, s+1)} \Biggl [ \sum_{h=1}^{2t} \bigl ( \frac{ (s+1)^3}{n}\bigr )^h \Biggr ] G_{m, c} (k,s+1-i)$$
and second for $t\le i\le 2t-1$ 
$$N_{m,s}  (i) \ p^{s+1-i} \le 
2 (s+1)^3 \sum_{k=1}^{\min (m, s+1)} \Biggl [ \sum_{h=0}^{2t} \bigl ( \frac{ (s+1)^3}{n}\bigr )^h \Biggr ] G_{m, c} (k,s+1-i).$$
At last, in using (\ref{minMaj}) with $s+1=\lfloor\hat{\gamma} \ln n \rfloor$ and with our choice for $\delta$ we obtain 
$$\sum_{i=1}^{t-1} N_{\lfloor \alpha \ln n \rfloor,s}  (i) \Bigl ( \frac{c}{4mn} \Bigr )^{s+1-i} \le D_1(\ln n )^{\frac{15}{2}}  \   n^{\alpha H(c) -2}$$
$$\sum_{i=t}^{2t-1} N_{\lfloor \alpha \ln n \rfloor,s}  (i) \Bigl ( \frac{c}{4mn} \Bigr )^{s+1-i} \le D_2(\ln n )^{\frac{9}{2}}  \   n^{\alpha H(c) -1-\delta}.$$

\end{proof}

\section{Proof of Proposition \ref{prop:function g}}\label{sec:technical}

Let us recall that  for any $1<c<2$ and  $  \alpha > 0$,  we consider the  domain   
 $\displaystyle \mathcal{D}_{\alpha} = \{ (\beta,
 \gamma) \mid  0< \beta \le \alpha  \hbox { and } \beta \le \gamma\}$ for the function $g_{\alpha,c}$ given from (\ref{eqn:def-of-g}) by  
\begin{equation}\label{eqn-of-g} 
 g_{\alpha,c}(\beta, \gamma)=  -1 +\alpha\ln\alpha-(\alpha-\beta)\ln(\alpha-\beta)+ \gamma\ln \Bigl [\frac{c \gamma}{2ex_0\alpha}\Bigr ]+ \beta\ln \Bigl [\frac{2(e^{x_0}-1)}{\beta} \Bigr ]
   \end{equation}
   \begin{equation}\label{eqn-of-g-diag} 
 g_{\alpha,c}(\beta, \beta)=  -1 +\alpha\ln\alpha-(\alpha-\beta)\ln(\alpha-\beta)+ \beta\ln \Bigl [\frac{c}{e \alpha} \Bigr ]
   \end{equation}

 with $x_0$  defined implicitly  when $0<\beta<\gamma$ by 
 \begin{equation}\label{x0}
 1-e^{-x_0}=\frac{\beta}{\gamma} x_0
 \end{equation}
 In the sequel, we shall write $g$ for $g_{\alpha,c}$ and $\mathcal{D}$ for $ \mathcal{D}_{\alpha} $.\\
 
  Proposition \ref{prop:function g} tells us that $g$ has a  strict and global maximum on $\mathcal{D}$  which is equal to $ \alpha H(c)-1
$ with
$\displaystyle H(c)=\ln c+ \Bigl( \frac{2}{c}-1\Bigr ) \ln(2-c).$ The proof of   Proposition \ref{prop:function g}  follows from  the  following claim :

\begin{claim}
\label{onevariableproof} For any $1<c<2$ and  $\alpha>0$,
 
 \begin{enumerate}
\item  for every fixed $\beta$ with $0<\beta \le  \alpha$, the function 
$\gamma \mapsto g(\beta, \gamma)$ is strictly concave on $[\beta, +\infty[$ with a strict  maximum at $\displaystyle\gamma_{\beta}=\frac{2\alpha}{c}\ln \Bigl ( \frac{ 2\alpha}{2\alpha - \beta c}\Bigr ).$

\item the function $\beta \mapsto g(\beta, \gamma_{\beta})$  is strictly concave on $]0,\alpha]$ with a  maximum at  $\displaystyle \hat\beta =  \frac{2\alpha(c-1)}{c}$, then  with $\displaystyle 
 \hat\gamma :=\gamma_{\hat\beta}= \frac{-2\alpha\ln (2-c)}{c},    g  (\hat\beta,  \hat \gamma  ) =\alpha H(c)-1.$
     
\end{enumerate}
\end{claim}

\begin{proof} 
 For the first point of this claim we compute, from (\ref{eqn-of-g}) and (\ref{x0}), the partial derivatives of $g$ with respect to $\gamma$. We get
  \begin{equation}\label{partialderivatives}
  \frac{ \partial g}{\partial\gamma}(\beta, \gamma) =
    \ln\left ( \frac{c\gamma}{2x_0\alpha}\right )\quad  \hbox{ and  } \quad  \frac{ \partial^2 g}{\partial\gamma^2} (\beta, \gamma)=
     \frac{\gamma-\beta x_0}{\gamma(\gamma - \beta(x_0+1))}.
     \end{equation}
      With  (\ref{x0}) we  first   observe that  
      \begin{equation}\label{eqn:first_ineq} \gamma-\beta
x_0=\gamma e^{-x_0}>0.
\end{equation}
Then
\begin{eqnarray*}
  \gamma-\beta (x_0+1) &=& \gamma-\beta x_0 - \beta \\
 \  &=& \gamma e^{-x_0} -\beta \\
  \  &=&  \gamma e^{-x_0} - \frac{\gamma (1-e^{-x_0})}{x_0} \\
  \  &=& \frac{\gamma}{x_0}(x_0e^{-x_0} -1+ e^{-x_0})
\end{eqnarray*}
let $\varphi(x)= xe^{-x} -1+ e^{-x}$. The function $\varphi$ is
decreasing with $\varphi(0)=0$. Hence, $\varphi(x_0)<0$ and
\begin{equation}\label{eqn:second_ineq}
     \gamma-\beta (x_0+1) <0.
\end{equation}
From the second identity in (\ref{partialderivatives}), (\ref{eqn:first_ineq}) and (\ref{eqn:second_ineq}) we conclude that $\displaystyle \frac{ \partial^2
g}{\partial\gamma^2} (\beta, \gamma)<0$. The  strict concavity of $\mapsto g(\beta, \gamma)$ follows.
Then the first identity in (\ref{partialderivatives}) and  (\ref{x0}) give the expected formula for the unique extremum, indeed we obtain 
 \begin{equation}\label{gammabeta}
     \gamma_{\beta} =\frac{2x_0 \alpha}{c}= \frac{2\alpha}{c}\ln \Bigl ( \frac{ 2\alpha}{2\alpha - \beta c}\Bigr )\hbox{ and } e^{x_0} -1 = \frac{\beta c}{2\alpha - \beta c}.
\end{equation} 

\medskip

 For the second point of the claim, observe that with  (\ref{eqn-of-g}) we have :
$$g(\beta, \gamma)=  -1 +\gamma\ln \Bigl [\frac{c \gamma}{2x_0\alpha}\Bigr ] -\gamma + 
 \alpha\ln\alpha-(\alpha-\beta)\ln(\alpha-\beta)+ 
 \beta\ln \frac{2(e^{x_0}-1)}{\beta}\;,$$
 thus from (\ref{gammabeta}) we obtain 
 \begin{equation}\label{g-in-gammabeta}
 g(\beta, \gamma_{\beta})= -1+ \alpha  \  K_c \Bigl ( \frac{\beta}{\alpha} \Bigr ) 
 \end{equation} 
 where for any  $x\in ]0,1[$, $\displaystyle K_c(x)= x\ln c + \Bigl (\frac{2}{c} -x  \Bigr ) \ln \Bigl ( 1- \frac{cx}{2} \Bigr ) -  (1 -x ) \ln  ( 1- x)$.
  $K_c$ is strictly concave on $]0,1[$ and reaches its maximum at $\displaystyle x= \frac{2(c-1)}{c}$. From (\ref{g-in-gammabeta}) with $\displaystyle \frac{\hat\beta}{\alpha} =  \frac{2(c-1)}{c}$ we get 
$\displaystyle\max_{\beta>0} g(\beta,\gamma_{\beta}) = -1+ \alpha  \  K_c \Bigl ( \frac{\hat\beta}{\alpha} \Bigr )= -1+ \alpha H(c)$. Then, with (\ref{gammabeta}) we obtain $\displaystyle \gamma_{\hat\beta}= \frac{2\alpha}{c}\ln \Bigl ( \frac{ 2\alpha}{2\alpha - \hat\beta c}\Bigr )= \frac{-2\alpha\ln (2-c)}{c}:=\hat\gamma\ .$  

At last, observe that  $\displaystyle \frac{ \partial g}{\partial\beta} (\beta, \gamma)= \ln\left (\frac{2(e^{x_0}-1)(\alpha-\beta)}{\beta} \right )$, so 
       $\hat\beta$ and  $
 \hat\gamma$ give the coordinates of the unique stationarity point of $g$, that is the unique solution of
  $\displaystyle  \frac{ \partial g}{\partial\beta}(\beta, \gamma)=
  \frac{ \partial g}{\partial\gamma}(\beta, \gamma)=0$.
 \\

\end{proof}

 \section{Conclusion}
 

We have performed an extensive study of a natural and expressive
quantified problem, $\onetwoqsat$. We have proved the existence of
a sharp phase transition from satisfiability to unsatisfiability
for $\onetwoqcnf$-formulas and we have given the exact location of the
threshold. The obtained results have several interesting features.
The  parameter  $m$, which is the number of universal variables,
controls the worst-case computational complexity of the problem
(which is ranging from linear time solvable to $\conp$-complete),
as well as the typical behavior of random instances.    When $m$
is small, there is a sharp threshold at $c=2$. On the other side,
when $m$ is large enough, actually   when $m>> \ln n$, there is a
sharp threshold at $c=1$: the analysis is similar, and in fact
easier, to what we have done for pure snakes in Section
\ref{sec:main}, in considering snakes with strictly distinct
universal variables, as shown in \cite{CreignouDER-08}. This fact should be compared to the fact
that the threshold location $c^*(\alpha)$ for $m=\lfloor \alpha\ln
n\rfloor$ goes to 1 when $\alpha$ goes to infinity. More
importantly, an original regime is observed when $m=\lfloor
\alpha\ln n\rfloor$. Using counting arguments on pure bicycles,
which are the seed of unsatisfiability, and on pure snakes, which
are special minimally false formulas, we got respectively a lower
and an upper bound for the threshold. It turns out that these two
bounds coincide, thus giving the exact location of the threshold
as a function of $\alpha$. 

A challenging question would be to determine the scaling window around
$c^*(\alpha)$ and get precise information on the typical contradictory
cycles that occur in random formulas inside this window.  

\end{document}